\documentclass[12pt]{iopart}
\usepackage{amssymb}
\usepackage{iopams}
\begin{document}
\title[Nonlinear Integral-Equation Formulation of Orthogonal Polynomials]
{Nonlinear Integral-Equation Formulation of Orthogonal Polynomials}
\author[Bender and Ben-Naim]{Carl~M.~Bender\footnote{Permanent address:
Department of Physics, Washington University, St. Louis MO 63130, USA} and
E.~Ben-Naim}
\address{Theoretical Division and Center for Nonlinear Studies, Los Alamos
National Laboratory, Los Alamos, NM 87545, USA}
\begin{abstract}
The nonlinear integral equation $P(x)=\int_\alpha^\beta
dy\,w(y)\,P(y)\,P(x+y)$ is investigated. It is shown that for a given
function $w(x)$ the equation admits an infinite set of polynomial
solutions $P(x)$. For polynomial solutions, this nonlinear integral
equation reduces to a finite set of coupled linear algebraic equations
for the coefficients of the polynomials. Interestingly, the set of
polynomial solutions is orthogonal with respect to the measure
$x\,w(x)$.  The nonlinear integral equation can be used to specify all
orthogonal polynomials in a simple and compact way. This integral
equation provides a natural vehicle for extending the theory of
orthogonal polynomials into the complex domain. Generalizations of the
integral equation are discussed.
\end{abstract}
\pacs{2.30.Rz, 2.10.Yn, 2.10.Ud}

There are many ways to specify uniquely a set of orthogonal
polynomials. One can specify the domain $(\alpha,\beta)$ and the
measure with respect to which the polynomials are orthogonal and then
use the cumbersome Gramm-Schmidt orthogonalization procedure to
construct the polynomials. For example, for the domain $(-1,1)$ and
measure $(1-x^2)^{-1/2}$, the Gramm-Schmidt procedure yields the
Chebyshev polynomials $T_n(x)$. Alternatively, one can specify a
recursion relation. The linear recursion relation
$T_{n+1}(x)=2xT_n(x)-T_{n-1}(x)$ along with the initial conditions
$T_0(x)=1$ and $T_1(x)=x$ again produces the Chebyshev
polynomials. Another approach is to give the differential-equation
eigenvalue problem satisfied by the polynomials. The Chebyshev
polynomials $T_n(x)$ satisfy the eigenvalue equation
$(1-x^2)y''(x)-xy'(x)+n^2y(x)=0$.  Stating the generating function is
yet another way to specify a set of polynomials. For the Chebyshev
polynomials the generating function $$\frac{1-xt}
{1-2xt+t^2}=\sum_0^\infty T_n(x)t^n$$ uniquely defines these
polynomials.

In this paper we propose a simple and compact way to specify a set of
orthogonal polynomials in terms of a nonlinear integral
equation. Consider the nonlinear integral equation
\begin{equation}
\label{int-eq}
P(x)=\int_\alpha^\beta dy\,w(y)\,P(y)\,P(x+y).
\end{equation}
The integration limits $\alpha$ and $\beta$ as well as the function
$w(x)$ are arbitrary except for the restriction that $w(x)$ must have
a nonvanishing integral, $\int_\alpha^\beta dx\,w(x)\neq0$. Therefore,
without loss of generality, we may assume that $w$ is normalized:
$$\int_\alpha^\beta dx\,w(x)=1.$$ Note that if we choose
$\alpha=-\infty$, $\beta=\infty$, and $w(y)=e^{iy}$, (\ref{int-eq})
reduces to the heavily studied equation for the Wigner function
\cite{cz,cuz}. The original motivation for considering this nonlinear
integral equation is that it describes a stochastic process in which
two random variables are subtracted to create a new one. The
steady-state probability distribution for this random variable
satisfies the integral equation (\ref{int-eq}) with $\alpha=0$,
$\beta=\infty$, and $w(y)=2$.

Even though this integral equation is nonlinear, its polynomial
solutions can be found analytically. By inspection one can see that
there is the trivial solution $P_0(x)=1$. However, there are also
infinitely many other polynomial solutions because (\ref{int-eq}) has
two remarkable properties.

First, if we seek a solution in the form of an arbitrary polynomial of degree
$n$,
\begin{equation}
\label{poly}
P_n(x)=\sum_{k=0}^na_{n,k}\,x^k,
\end{equation}
the nonlinear integral (\ref{int-eq}) preserves the degree of the
polynomial.  For example, if we substitute an arbitrary linear
polynomial $P_1(x)=a_{1,0}+ a_{1,1}x$ or an arbitrary quadratic
polynomial $P_2(x)=a_{2,0}+a_{2,1}x+a_{2,2} x^2$ into (\ref{int-eq}),
we obtain
$$a_{1,0}+a_{1,1}x=\langle
P_1(y)\left[a_{1,0}+a_{1,1}y+a_{1,1}x\right]\rangle$$ and
$$a_{2,0}+a_{2,1}x+a_{2,2}x^2=\langle P_2(y)\left[a_{2,0}+a_{2,1}y+a_{2,1}x+
a_{2,2}y^2+2a_{2,2}xy+a_{2,2}y^2\right]\rangle,$$
where we have introduced the notation
\begin{equation}
\langle f\rangle\equiv\int_\alpha^\beta dx\,w(x)f(x).
\end{equation}
In general, for an arbitrary polynomial of degree $n$ both the left
and right sides of (\ref{int-eq}) are polynomials of the same degree
$n$.

Substituting a polynomial of degree $n$ into the right side of
(\ref{int-eq}) and performing the integration, we obtain a polynomial
of degree $n$.

Second, for the case of polynomial solutions, the nonlinear equation
(\ref{int-eq}) reduces to a system of coupled {\it linear} equations,
which are obtained by equating like powers of $x$. For the case $n=1$
above, equating the coefficient of $x^1$ on both sides of the equation
gives $a_{1,1}=a_{1,1}\langle P_1(y)\rangle$. We require that the
polynomial have degree one, $a_{1,1}\neq 0$, so
\begin{equation}
\label{p0av}
\langle P_1(x)\rangle=1,
\end{equation}
where we have replaced $y$ by $x$.  Next, we equate the coefficients
of $x^0$ and get \hbox{$a_{1,0}=a_{1,0}\langle
P_1(y)\rangle+a_{1,1}\langle y\,P_1(y)\rangle$}. Substituting
(\ref{p0av}) into this equation, we obtain
\begin{equation}
\label{p1av}
\langle x\,P_1(x)\rangle=0.
\end{equation}
Thus, the nonlinear equation (\ref{int-eq}) reduces to a system of two
coupled inhomogeneous linear equations for the two unknowns $a_{1,0}$
and $a_{1,1}$.  The integral equation reduces to a linear algebraic
system only for polynomials because, as we see in the above
calculation, the derivation relies on the fact that the power series
in $x$ truncates at a finite order.

Let us repeat this procedure for quadratic polynomials: Starting at
the highest power $x^2$, we find that $a_{2,2}=a_{2,2}\langle
P_2(y)\rangle$ and hence we recover (\ref{p0av}) with the subscript 1
replaced by 2. Setting the coefficients of $x^1$ equal gives
$a_{2,1}=a_{2,1}\langle P_2(y)\rangle+2a_{2,2} \langle
y\,P_2(y)\rangle$ and we recover (\ref{p1av}) with the subscript 1
replaced by 2. The coefficient of $x^0$ gives $a_{2,0}=a_{2,0}\langle
P_2(y) \rangle+a_{2,1}\langle y\,P_2(y)\rangle+a_{2,2}\langle
y^2\,P_2(y)\rangle$.  Therefore, there is now the third equation
\begin{equation}
\label{p2av}
\langle x^2\,P_2(x)\rangle=0.
\end{equation}

The general pattern is clear. An $n$th-degree polynomial is a solution
of the integral equation if and only if the following set of $n+1$
linear equations is satisfied:
\begin{equation}
\label{linear}
\langle x^kP_n(x)\rangle=\delta_{k,0}\qquad (k=0,1,\ldots,n).
\end{equation}

The linear equations (\ref{linear}) imply that the
polynomials $P_n(x)$ are orthogonal with respect to the measure
\begin{equation}
\label{measure}
g(x)=x\,w(x).
\end{equation}
If $P_m(x)=\sum_{k=0}^m a_{m,k}\,x^k$ is a polynomial of degree $m<n$,
then from (\ref{linear}) we conclude that
\begin{equation}
\label{orthog}
\langle x\,P_nP_m\rangle=\sum_{k=0}^{m}a_{m,k}\langle x^{k+1}\,P_n(x)\rangle=
\sum_{k=1}^{m+1}a_{m,k-1}\langle x^k\,P_n(x)\rangle=0.
\end{equation}
This is the main result of our paper. We have shown that the nonlinear
integral equation (\ref{int-eq}) admits an infinite set of polynomial
solutions. This set is unique; there is one and only one polynomial
solution of degree $n$. These polynomials are orthogonal.

We comment that this integral formulation of orthogonal polynomials is general.
The integration limits and the function $w(x)$ are arbitrary apart from the
restriction that $w(x)$ have a positive integral $\int_\alpha^\beta\,dx\,w(x)>
0$.

We mention three classical orthogonal polynomials \cite{as} that can be
generated using this approach:

$\bullet$ Generalized Laguerre polynomials $L_n^{(\gamma)}(x)$ using
$\alpha=0$, $\beta=\infty$, and $w(x)=x^{\gamma-1}e^{-x}$ for all
$\gamma\geq 1$;

$\bullet$ Jacobi polynomials $G_n(p,q,x)$ using $\alpha=0$, $\beta=1$,
and $w(x) =x^{q-2}(1-x)^{p-q}$, for $q>1$ and $p-q>-1$;

$\bullet$ Shifted Chebyshev polynomials of the second kind $U_n^*(x)$ using
$\alpha=0$, $\beta=1$, and $w(x)=(1-x)^{1/2}x^{-1/2}$.

The equations (\ref{linear}) can be written compactly as $n+1$
simultaneous linear equations for the $n+1$ coefficients $a_{n,j}$:
\begin{equation}
\sum_{j=0}^n a_{n,j}m_{k+j}=\delta_{k,0}\qquad(k=0,1,\ldots,n).
\end{equation}
Here $m_n=\langle x^n\rangle$ are the moments of $w(x)$ and by
assumption $m_0= 1$. Thus, using Cramer's rule we can express the
polynomial solutions explicitly as ratios of determinants
\cite{aar}. We define the matrices
$$A_n=\left(\begin{array}{cccc} 1&x&\cdots &x^n\\ m_1&m_2&\cdots
&m_{n+1}\\ \vdots&\vdots&\ddots &\vdots\\ m_n&m_{m+1}&\cdots & m_{2n}
\end{array}\right),\quad B_n=\left(\begin{array}{cccc} m_0&m_1&\cdots
&m_n\\ m_1&m_2&\cdots &m_{n+1}\\ \vdots&\vdots&\ddots &\vdots\\
m_n&m_{m+1}&\cdots & m_{2n} \end{array}\right),$$ so that $B_n=\langle
A_n\rangle$. The polynomials $P_n(x)$ are then given by
\begin{equation}
P_n(x)=\frac{\det A_n}{\det B_n}.
\label{pee}
\end{equation}

The normalization of the polynomials can be given in the form
\begin{equation}
\langle x\,P_n(x)\,P_m(x)\rangle=\delta_{n,m}\,G_n.
\end{equation}
The normalization factors $G_n$ are ratios of determinants of matrices
\begin{equation}
G_n=\frac{(\det C_n)(\det C_{n+1})}{(\det B_{n+1})^2},
\label{g-whiz}
\end{equation}
where the matrix $C_n$ is given by
$$C_n=\left(\begin{array}{cccc} m_1&m_2&\cdots &m_n\\
m_2&m_3&\cdots &m_{n+1}\\ \vdots&\vdots&\ddots &\vdots\\
m_n&m_{m+1}&\cdots & m_{2n} \end{array}\right).$$

For completeness, we give the first three polynomials explicitly:
\begin{eqnarray*}
\label{lowest-p}
P_0(x)&=&1,\\
P_1(x)&=&\frac{m_2-xm_1}{m_2-m_1^2},\\
P_2(x)&=&\frac{(m_2m_4\!-\!m_3^2)+(m_2m_3\!-\!m_1m_4)x+(m_1m_3\!-\!m_2^2)x^2}
{m_4(m_2-m_1^2)-m_3^2+2m_1m_2m_3-m_2^3}.
\end{eqnarray*}
Also, we give the corresponding normalization factors $G_n$:
\begin{eqnarray*}
\label{lowest-g}
G_0&\!=\!&m_1,\\ G_1&\!=\!&m_1\frac{m_1m_3-m_2^2}{(m_2-m_1^2)^2},\\
G_2&\!=\!&\!\frac{(m_1m_3\!-\!m_2^2)(m_1m_3m_5\!-\!m_2^2m_5\!-\!m_1m_4^2\!
+\!2m_2m_3m_4\!-\!m_3^3)}{[m_4(m_2-m_1^2)-m_3^2+2m_1m_2m_3-m_2^3]^2}.
\end{eqnarray*}
Note that (\ref{pee}--\ref{g-whiz}) are only valid if $\det B_n\neq0$. This
condition holds when the measure $g(x)=x\,w(x)$ is positive on $\alpha\leq
x\leq\beta$ \cite{baker,bmp}.

The integral equation (\ref{int-eq}) specifies all possible orthogonal
polynomials. When the weight function $g(x)$ with respect to which the
polynomials are orthogonal does not vanish at $x=0$ and consequently
$w(x)=g(x)/x$ is singular at $x=0$, the path of integration from
$\alpha$ to $\beta$ may be taken in the complex plane to avoid the
singularity at the origin. Using a complex integration path, all steps
leading to (\ref{orthog}) are valid.

For example, consider the Legendre polynomials for which $\alpha=-1$,
$\beta=1$, and $g(x)$ is a constant. There are an infinite number of
topologically distinct integration paths that connect $-1$ to
$1$. These paths are characterized by their winding numbers. For
definiteness, we choose a path that goes from $-1$ to $1$ in the
positive (counterclockwise) direction and does not encircle the
origin. On this path $\int dx/x=i \pi$ and hence to maintain the
normalization $\int_\alpha^\beta dx\,w(x)=1$ we use $w(x)=1/(i\pi
x)$. The moments $m_n=\langle x^n\rangle$ are $m_0=1$, $m_1=2/(i\pi)$,
$m_2=0$, $m_3=2/(3i\pi)$, $\ldots$\,\,. From the moment formulas
(\ref{pee}), we obtain
\begin{eqnarray}
P_0(x)&=&1,\nonumber\\
P_1(x)&=&\frac{i\pi}{2}x,\nonumber\\
P_2(x)&=&1-3x^2,\nonumber\\
P_3(x)&=&\frac{3i\pi}{8}(3x-5x^3),
\label{whoopie}
\end{eqnarray}
and so on. These polynomials are the standard Legendre polynomials,
except that the odd polynomials have an imaginary multiplicative
factor that is determined by the winding number of the integration
path. Of course, these polynomials are solutions of the linear
equations (\ref{linear}).

To summarize, while the usual theory of orthogonal polynomials is
formulated in terms of real integrals, the integral equation
(\ref{int-eq}) provides a simple and natural framework to extend the
theory of orthogonal polynomials into the complex domain. In doing so
we discover an interesting connection between the polynomial
coefficients and the topological winding number of the integration
path.

There are many ways to generalize the integral equation (\ref{int-eq}):

\noindent{\bf 1.~Multiplicative argument.}
If we replace the term $P(x+y)$ in the original nonlinear integral equation
(\ref{int-eq}) by $P(xy)$, we obtain a new class of nonlinear equations:
\begin{equation}
\label{int-eq-mult}
P(x)=\int_\alpha^\beta dy\,w(y)\,P(y)\,P(x\,y).
\end{equation}
Each of these nonlinear integral equations also has an infinite number of
polynomial solutions.

Again, there is the constant solution $P_0(x)=1$. To find other
solutions we substitute a polynomial of degree $n$ and then equate
coefficients of $x^k$ on the left and right sides, starting with
$k=n$. The nonlinear integral equation reduces to a set of $n+1$
inhomogeneous equations of the form
\begin{equation}
a_{n,k}\langle x^kP_n(x)\rangle=a_{n,k}\qquad(k=0,1,\ldots,n).
\end{equation}
However, unlike the previous case, the equations are quadratic and
there are now $2^{n-1}$ solutions because each of the coefficients
$a_{n,k}$ can be either zero or nonzero for $k=0,\ldots,n-1$. For each
nonzero coefficient the linear equation
\begin{equation}
\langle x^kP_n(x)\rangle=1
\label{uuu}
\end{equation}
holds for all $k$ for which $a_{n,k}\neq0$.

In one special class of solutions all of the coefficients are nonzero and the
polynomials are orthogonal with respect to the measure
\begin{equation}
g(x)=(1-x)\,w(x).
\label{lkjf}
\end{equation}
To verify this, we take a polynomial of degree $m<n$ and observe that
$$\langle (1-x)P_nP_m \rangle=\sum_{k=0}^m a_{m,k} \left(\langle
x^kP_n\rangle-\langle x^{k+1}P_n\rangle\right)=0.$$ This equation is
valid because all of the terms in the parentheses vanish by virtue of
(\ref{uuu}). This measure is relevant for the class of polynomials for
which $0\leq\alpha<\beta\leq1$.

Here are two classical orthogonal polynomials that can be generated in
this way:

$\bullet$ Jacobi polynomials $G_n(p,q,x)$ using $\alpha=0$, $\beta=1$,
and $w(x) =(1-x)^{p-q-1}x^{q-1}$ with $p-q>0$ and $q>0$;

$\bullet$ Shifted Chebyshev polynomials of the second kind $U_n^*(x)$ using
$\alpha=0$, $\beta=1$, and $w(x)=x^{1/2}(1-x)^{-1/2}$.

In another class of solutions, the coefficients alternate between zero
and nonzero so that the polynomials alternate between definite even
and odd parity.  The polynomials are orthogonal with respect to a
measure whose moments $\mu_n$ are
\begin{equation}
\label{mn-even}
\mu_n=\textstyle{\frac{1}{2}}\langle x^n-x^{n+2}\rangle\left[1+(-1)^n\right].
\end{equation}
Thus, all the even moments of the measure are positive and all the odd
moments vanish. It is easy to show that polynomials of similar parity
are orthogonal with respect to the measure
$(1-x^2)\,w(x)$. Polynomials of dissimilar parity are orthogonal
because their product is an odd polynomial, and the odd moments vanish
(\ref{mn-even}).

\noindent{\bf 2.~Linear arguments.}
For the integral equation with a linearly shifted argument,
\begin{equation}
P(x)=\int_\alpha^\beta dy\,w(y)\,P(y)\,P(x+a+b\,y),
\end{equation}
where $b\neq0$ is an arbitrary constant, polynomials of degree $n$ are
solutions when $\langle (a+bx)^k\,P_n(x)\rangle=\delta_{k,0}$. This
implies that the polynomials are orthogonal with respect to the
measure $g(x)=(a+bx)\,w(x)$.  Note that when $a=0$, the polynomials
are identical to those generated by the original integral formula
(\ref{int-eq}). Curiously, for $a=1$ and $b=-1$ we recover the
polynomials generated by the integral equation (\ref{int-eq-mult}).

\noindent{\bf 3.~Functional arguments.}
The integral equation
\begin{equation}
P(x)=\int_\alpha^\beta dy\,w(y)\,P(y)\,P[x+f(y)],
\label{eli}
\end{equation}
where $f(x)$ is a nonconstant function, has polynomial solutions of
degree $n$ when $\langle [f(x)]^k\,P_n(x)\rangle=\delta_{k,0}$. These
polynomials are not necessarily an orthogonal set. Nevertheless, the
polynomial $P_n(x)$ is orthogonal to the function $P_m[f(x)]$ when
$m<n$ with respect to the measure $g(x)=f(x)\,w(x)$.

\noindent{\bf 4.~Arbitrary functions.}
The integral equation
\begin{equation}
P(x)=\int_\alpha^\beta dy\,w(y)\,f[P(y)]\,P(x+y),
\end{equation}
where $f(x)$ is a nonconstant function, has polynomial solutions of
degree $n$ when $\langle x^k f[P_n(x)]\rangle=\delta_{k,0}$. This
implies that the function $f[P_n(x)]$ is orthogonal to the polynomial
$P_m(x)$ with respect to the measure $g(x)=x\,w(x)$.

\noindent{\bf 5.~Further generalizations.}  It is worth considering
what happens when the function $w(x)$ is singular on the interval
$\alpha<x<\beta$. Also, an obvious way to generalize (\ref{int-eq}) is
to iterate it, and thereby to obtain integral equations that are
cubic, quartic, and so on. Furthermore, one can study the properties
of nonpolynomial solutions to (\ref{int-eq}). We have found many such
solutions. Finally, one can generalize (\ref{int-eq}) to
multidimensional integrals and study the properties of the resulting
multivariate polynomial solutions.

In summary, we have shown that all sets of orthogonal polynomials are
solutions of nonlinear integral equations. For polynomial solutions
these nonlinear equations reduce to simultaneous linear equations for
the coefficients of the polynomials. The measure with respect to which
the polynomials are orthogonal depends on the form of the integral
equation and on the integration measure. The nonlinear integral
equations discussed here provide a simple and compact way to define a
set of orthogonal polynomials and also provide a framework for
extending the general theory of orthogonal polynomials into the
complex domain.

\vspace{0.5cm}
\begin{footnotesize}
\noindent
We thank R.~Askey, P.~L.~Krapivsky, R.~Laviolette, P.~Nevai, M.~Nieto,
and R.~Theodorescu for useful discussions. We acknowledge financial
support from the U.S. DOE grant DE-AC52-06NA25396.
\end{footnotesize}

\end{document}